\documentclass[aps,pre,superscriptaddress,reprint,
amsmath,amssymb,floatfix]{revtex4-1}
\usepackage{graphicx}
\usepackage{bm}
\usepackage[colorlinks]{hyperref}
\usepackage{siunitx}

\newcommand{\AVE}[1]{\ensuremath{\left\langle {#1} \right\rangle}}
\newcommand{\ABS}[1]{\ensuremath{\left\lvert {#1} \right\vert}}

\newcommand{\bF}{\ensuremath{\bm{F}}}

\newcommand{\bS}{\ensuremath{\bm{S}}}

\newcommand{\bU}{\ensuremath{\bm{U}}}

\newcommand{\bj}{\ensuremath{\bm{j}}}

\newcommand{\bn}{\ensuremath{\bm{n}}}

\newcommand{\bx}{\ensuremath{\bm{x}}}

\newcommand{\Da}{\ensuremath{\mathrm{Da}}}
\newcommand{\Pe}{\ensuremath{\mathrm{Pe}}}

\begin{document}
	
	
	\title{Anti-Swarming: Structure and Dynamics of Repulsive Chemically Active
		Particles}
	
	
	
	\author{Wen Yan}
	\email[]{wyan@flatironinstitute.org}
	\affiliation{Department of Mechanical \& Civil Engineering, 
		Division of Engineering \& Applied Science, 
		California Institute of Technology}
	\altaffiliation{Current Address: Center for Computational Biology, Flatiron Institute, Simons
		Foundation}
	\author{John F. Brady}
	\email[]{jfbrady@caltech.edu}
	\affiliation{Division of Chemistry \& Chemical Engineering 
		and Division of Engineering \& Applied Science 
		California Institute of Technology}
	

	
	\date{\today}
	
	\begin{abstract}
		Chemically active Brownian particles with surface catalytic reactions may repel
		each other due to diffusiophoretic interactions  in the reaction and product
		concentration fields. The system behavior can be described by a `chemical'
		coupling parameter $\Gamma_c$ that compares the strength of diffusiophoretic
		repulsion to Brownian motion, and by a mapping to the classical electrostatic
		One Component Plasma (OCP) system. When confined to a constant-volume domain,
		Body-Centered Cubic crystals  spontaneously form from random initial
		configurations when the repulsion is strong enough to overcome  Brownian motion.
		Face-Centered Cubic crystals may also be stable. The `melting point' of the
		`liquid-to-crystal transition' occurs at $\Gamma_c\approx140$ for both BCC and
		FCC lattices.
	\end{abstract}
	
	
	\maketitle
	
	\section{Introduction}
	Chemically active particles suspended in fluids may achieve self-propulsion by
	surface catalytic reactions of chemical solutes \cite{EbbensReview2010}.
	One mechanism is self-diffusiophoresis, whereby the motion of a particle arises from
	the asymmetric solute concentration field $c(\bx,t)$ created near its surface.
	Typically, reactants are consumed on the surface of a chemically active particle,
	and when a second particle appears is in the vicinity, it is attracted by a
	diffusiophoretic velocity $\bU \sim - \nabla c$. Active particles with
	attractive interactions are observed to exhibit dynamic clustering and
	gas-liquid phase transition \cite{Theurkauff2012,Buttinoni2013,Palacci2013}.
	Thermodynamic-like theories \cite{ThermoAct2015} utilizing the swim pressure
	\cite{Pressure2014} as an equation of state, and other theories based on similar
	thermodynamic-like models \cite{Stenhammar2013,Solon2015a,Cates2015} work well
	in describing the phase separation phenomena.
	
	However, few studies have investigated active particles with repulsive
	interactions. If the surface chemical reactions release solutes instead of
	consuming them, the solute concentration $c(\bx,t)$ is increased in the vicinity
	of each particle, and the diffusiophoretic velocity is now repulsive between
	particles (c.g Fig.~\ref{fig:schematic}). Repulsive particles, if confined in a
	constant volume container, may overcome the randomizing thermal Brownian motion
	and form a crystal lattice \cite{Soh2008}.
	\citet{Derjaguin1984} observed the formation of periodic crystal-like structures
	in living cells and suggested that it is due to  repulsive diffusiophoretic
	interactions.
	
	A classical example of repulsive particles that show a liquid-to-crystal
	transition is the so-called One Component Plasma (OCP). In an OCP moving
	positive charges are immersed in a uniform and neutralizing background sea of
	negative electrons, and the system behavior is governed by an electrostatic
	coupling parameter $\Gamma_e$, which measures the electrostatic energy relative
	to thermal energy \cite{Brush1966}. It is well known that the liquid-like
	structure at small $\Gamma_e$ transforms to BCC (body-centered cubic) for
	$\Gamma_e\gtrsim 175$
	\cite{Gillan1974,Rogers1974,Stroud1976,Itoh1977,Bernu1979,Baus1980,Tan1995,DeWitt2001,Chugunov2003,Daligault2006}.
	
	In this work we explore the collective motion of repulsive active particles by
	simulations with a full solution of the diffusiophoretic interactions as
	described in our methods paper \cite{yan_method_2016}. We show that repulsive
	chemically active particles exhibits a `liquid-to-crystal' phase transition,
	similar to an OCP. Quantitatively, we define a chemical coupling parameter
	$\Gamma_c$ for the chemically active system in analogy to $\Gamma_e$ for an OCP.
	
	\section{Problem formulation}
	
\begin{figure}[h]
		\includegraphics[width=\linewidth]{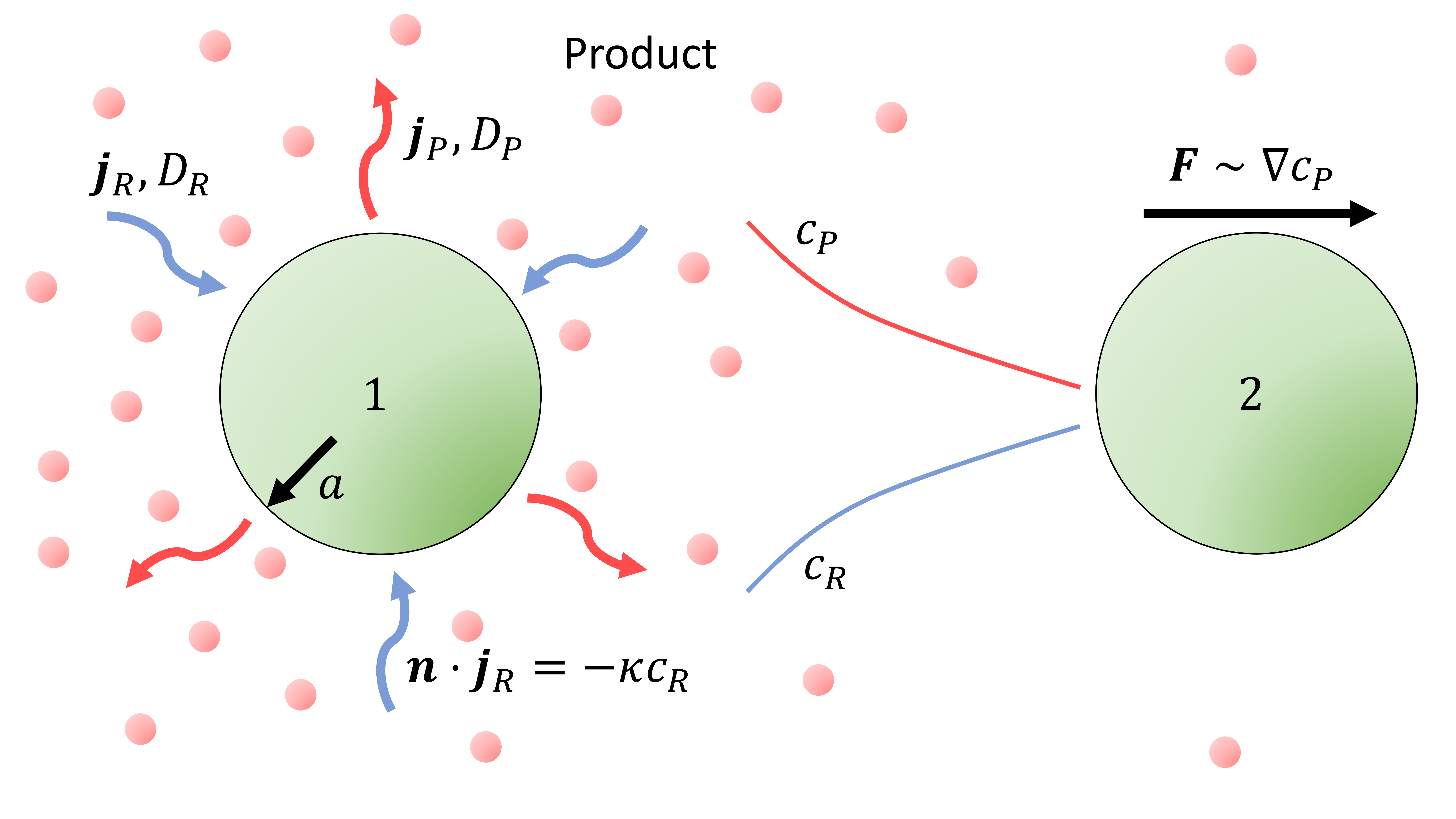}
		\caption{{ A schematic for the repulsive diffusiophoretic interaction. Reactant
				solutes with diffusivity $D_R$ and flux $\bj_R$ are consumed on the surface
				of 
				catalytic particle $1$ with the first-order boundary condition $\bn\cdot\bj_R
				=
				-\kappa c_R$. Product solutes (small red dots) leave the particle surface
				with
				flux $\bj_P$ and diffusivity $D_P$. In the vicinity of the particle $1$ $c_P$
				is
				increased while $c_R$ is decreased. When the product is more effective at
				pushing particle $2$ (which occurs when $1-\theta D_R/D_P < 0$)  particle $2$
				feels a repulsive diffusiophoretic force $\bF\sim\nabla c_P$.  From the
				reaction
				boundary conditions a simple scaling between $c_R$ and $c_P$ holds
				\cite{CordovaFigueroa2008} and only the reaction concentration field needs to
				be
				computed, which we denote simply as $c$.  The particles then move according
				to
				(\ref{eq:U0def}).  \label{fig:schematic}}}
	\end{figure}

	A first order surface catalytic reaction $R\rightarrow \theta P$ is assumed to
	occur homogeneously on the spherical particle surface as illustrated in
	Fig.~\ref{fig:schematic}. Making use of the stoichiometry/diffusivity factor
	$(1-\theta D_R/D_P)$, the reaction can be taken to be irreversible:
	$\bj_R\cdot\bn=- \kappa c(\bn)$ on the boundary, where $c$ is the reactant
	concentration, $\kappa$ is the reaction rate constant and $\bn$ is the surface
	normal vector pointing outward from the particle. Here, $\theta$ is the
	stoichiometry of the reaction and $D_R$ and $D_P$ are the diffusivities of the
	reactants and products, respectively. The Damkh\"{o}ler number $\Da=a\kappa/D_R
	$ governs the reaction rate: $\Da\to \infty$ corresponds to diffusion limited
	due to a fast reaction, while $\Da\to 0$ is the slow reaction-rate limit.  
	
	When the chemical solutes are much smaller in size than the active particles,
	each chemically active particle is driven by the osmotic pressure of the
	reactant solute concentration $k_BTc(\bx,t)$ integrated over the particle's
	surface \cite{CordovaFigueroa2008,Brady2011} and achieves the velocity
	\begin{equation}\label{eq:U0def}
		\bU_0 = -  (1-\theta D_R/D_P) \frac{L(\Delta)}{6\pi\eta a}  \oint \bn\,  k_BT
		c(\bx, t) dS \,,
	\end{equation}
	where $a$ is the particle radius, $\eta$ is the solution viscosity, and the
	nondimensional hydrodynamic mobility function $L(\Delta) = (3/2)\Delta^2(1 +
	{\textstyle{\frac{2}{3}}}\Delta)/(1 + \Delta)^3$, with $\Delta = \delta/a$,
	measures the flow of fluid with viscosity $\eta$ in a layer of thickness
	$\delta$ adjacent to the colloidal particle where the particle-solute
	interactive force is operative.
	Here we have taken the simplest form of interactive force between the solute and
	the colloidal particle, namely a hard-sphere repulsive force at a distance $r_c
	= a + \delta$ (and $\delta$ need not be small compared to the particle size $a$,
	although typically it is so).
	More general interactive forces will only have a quantitative effect and the
	details are discussed in \cite{Brady2011}.
	The prefactor $(1-\theta D_R/D_P)$ scales the solution of reactant concentration
	$c(\bx,t)$ to the total solute concentration of both reactant and products. When
	$\theta D_R/D_P>1$, the products push the particles more effectively than the
	reactants and the particles can be considered sources releasing products and
	therefore they repel each other.
	

	The governing equation for $c(\bx,t)$ is the classic
	convection-reaction-diffusion equation.
	The convection is controlled by the P\'{e}clet number $\Pe={U_0a}/{D_R}$.
	In diffusiophoresis, the particle velocity $U_0$ is usually so small that
	$\Pe\ll 1$  \cite{CordovaFigueroa2008}, and therefore the convection of $c$ can
	be ignored. Diffusion of the reactive solute is fast enough for $c$ to achieve a
	steady state, instantaneously following the particle motion.
	In this case, the governing equation for the reactant  reduces to Laplace's
	equation, $\nabla^2 c = 0$, similar to an electrostatic field. To leading order,
	the disturbance to the solute concentration field induced by one reactive
	particle is $c' \sim q/r$, where $q$ is the particle reactivity---that is, how
	many molecules are consumed on the particle surface in unit time, which is
	analogous to the electrostatic charge $Ze$.
	
	The active particles are assumed to be confined in a constant volume three
	dimensional space, and the reactant is assumed to be released by distributed
	sources throughout the space to maintain the system as `chemically neutral.'
	Therefore the volume average reactant concentration is maintained at a constant
	$\AVE{c}$. Without the chemically neutralizing condition, the particles
	eventually consume all the reactant and no steady state can be achieved.
	Experimentally, \citet{Theurkauff2012} have demonstrated a 2D implementation of
	a chemically neutral suspension in which the solute diffuses into a colloid
	monolayer reaction zone from a large reservoir and the system is kept evolving
	for many hours to reach a steady state. The chemically neutral assumption is
	also common for 3D reactive suspension systems \cite{Bonnecaze1991a}.
	
	An analogy to an OCP can be made.
	The repulsive active particles resemble the positive ions in an OCP, and the
	chemically neutralizing sources are similar to the electrostatically
	neutralizing background.
	By analogy, active particles should be liquid-like when the repulsion is weak
	and be solid-like when the repulsion is strong enough to order the particles
	into a periodic lattice. 
	
	A key difference, however, is that moving ions in an OCP are point charges and
	the charges are fixed at $Ze$, while the reactivity $q$ of a chemically active
	particle changes in response to the local concentration of reactants due to the
	chemical reaction on the particle's surface. Also, the reactivity has a
	distribution on the particle's spherical surface---the particle is more than
	merely a `point charge.' The changing reactivity results in changing
	interactions, which is fundamentally different from the additive pairwise
	potential assumption employed in previous simulation work on attractive active
	particles \cite{Palacci2013,Redner2013a}.
	The changing reactivity also poses a  difficulty for thermodynamic-like
	treatments.
	Even if we define a mean-field effective pairwise potential, it is
	state-dependent, and it is known that some thermodynamic inconsistencies and
	peculiarities may appear for density-dependent pairwise interactions
	\cite{Louis2002,Tejero2003}. 
	
	In this work, we simulate the system with the Accelerated Laplacian Dynamics
	method \cite{yan_method_2016}, which we  describe briefly without going into the
	mathematical details. The chemical reaction on each particle is represented by a
	multipole expansion, keeping only the monopole, $q$, and the dipole $\bS$,
	similar to electrostatics. Here, $q$ is the net consumption rate of reactant and
	$\bS$ is the asymmetry of the consumption on the particle surface. Second, the
	perturbation $c'$ of each particle to the average field $\AVE{c}$  is calculated
	from $q$, which propagates as $1/r$, and $\bS$, which propagates as $1/r^2$.
	Third, with the first order reaction condition, the monopole and dipole strength
	of particle $\alpha$ follow from a Faxen-type law:  $q_{\alpha}\propto
	\AVE{c}+c'(\bx_{\alpha})$, and  $\bS_{\alpha}\propto \nabla c'(\bx_{\alpha})$,
	where $c'(\bx_{\alpha})$ and $\nabla c'(\bx_{\alpha})$ are perturbations arising  from
	all particles $\beta\neq\alpha$ and are evaluated at the center of $\alpha$.
	(For chemically neutral systems $\nabla \AVE{c} = 0$.) In this way, the
	equations for the solute  field  $c$ are closed and can be solved iteratively at
	each timestep for different configurations of the active particles.
	
	The diffusiophoretic velocity of an active particle is then determined from the
	solution for the solute concentration field $c$ at each timestep. The velocity
	$\bU_{0,\alpha}$ of particle $\alpha$  in (\ref{eq:U0def}) can be calculated
	analytically utilizing the first-order reaction boundary condition,
	$\bj_R\cdot\bn=- \kappa c(\bn)$, to give
	\begin{equation}\label{eq:U0nonD}
		\frac{\bU_{0,\alpha}}{D/a} = - \left(1-\theta D_R/D_P\right)L(\Delta)\AVE{c}a^3
		\frac{4\pi a  \nabla c(\bx_{\alpha})}{ \left(Da+2\right)\AVE{c}} \,.
	\end{equation}
	By the assumed uniformity of the reaction on a particle surface  there is no
	self-diffusiophoretic motion; particle motion arises solely from normal
	diffusiophoresis in the concentration gradient created by the other particles. 
	
	The system dynamics are integrated with over-damped Brownian dynamics: $\Delta X
	= \bU_0 \Delta t  + \Delta X^B + \Delta X^{HS}$, where $\Delta X^B$ is the
	translational Brownian motion satisfying $\AVE{\Delta X^B}=0, \AVE{\Delta
		X^B\Delta X^B}=2D\Delta t$, and $\Delta X^{HS}$ is the non-overlapping
	hard-sphere collision displacement calculated with the potential-free algorithm
	\cite{Foss2000}.

	We nondimensionalize the system with the active particle radius $a$,  the
	particle diffusion time $\tau_D = a^2/D$, where $D$ is the Brownian diffusivity
	of an active particle, $D = k_BT/6\pi\eta a$, and the imposed reactant
	concentration $\AVE{c}$.
	The phoretic velocity then behaves as  $\bU_0\propto -S_D \nabla
	\left(c/\AVE{c}\right)$, where $S_D = \left(1-\theta
	D_R/D_P\right)L(\Delta)\AVE{c}a^3$ is the nondimensional concentration, i.e.,
	the `fuel concentration.' Increasing $\ABS{S_D}$ is equivalent to increasing the
	hydrogen peroxide concentration in the experiments
	\cite{Howse2007,Theurkauff2012}.
	In this work, we report only the result of the repulsive case $S_D<0$. The
	attractive case for $S_D>0$ is discussed elsewhere \cite{yan_dynamics_2016}.
	
	
	
	\section{The weak repulsion regime: fluctuating interactions}
	In simulations covering a wide range of volume fraction $0.001<\phi<0.15$ and
	Damk\"{o}hler number $0.1<\Da<10$, we found that under weak repulsion (small
	$\ABS{S_D}$), the system remains randomly distributed due to Brownian motion. To
	analyze the structure Voronoi cells are built around each particle and the local
	volume fraction is defined as $\phi_p = \frac{4}{3}\pi a^3 /V_p  $, where $V_p$
	is the volume of the Voronoi cell occupied by that particle.
	
	\begin{figure}[ht]
		\centering
		\includegraphics[width=\linewidth]{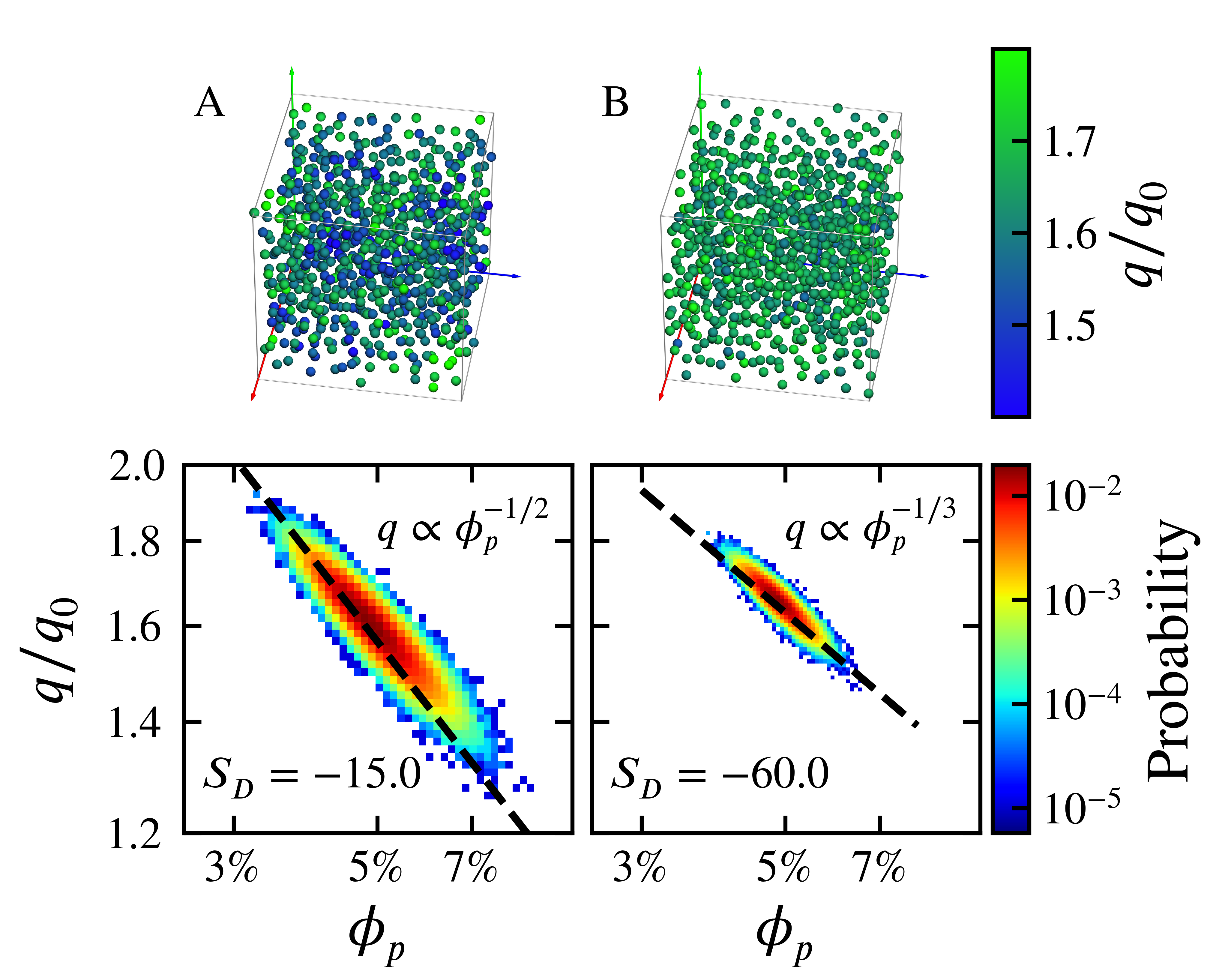}
		\caption{The distribution of particle reaction $q$ and local volume fraction
			$\phi_p$, and their correlation. A: The snapshot of the equilibrium structure of
			system in a periodic box of $42a\times42a\times42a$, with $\phi=0.0488$,
			$\Da=2.0$, $N=864$, $S_D = -15.0, \Gamma_c\approx 20$. Each particle is colored
			by $q/q_0$. B: The same system, but equilibrated with stronger repulsion $S_D =
			-60.0, \Gamma_c\approx 80$. }
		\label{fig:qphicorr}
	\end{figure}
	
	The first order reaction $R\to\theta P$ gives an infinitely dilute reactivity
	$q_0= - 4\pi  D_R a \AVE{c} \Da/(1+\Da)$. With increasing $\phi$, many-body
	interactions increase the reactivity of a particle and the average  reactivity
	$\AVE{q}$ increases \cite{Bonnecaze1991a}. In this work we assume that the
	volumetric average of reactant solute concentration is held constant at
	$\AVE{c}$ by the distributed source of reactant, and therefore on average
	$\AVE{q}/q_0 > 1$ as shown in Fig.~\ref{fig:qphicorr}. Although in principle
	$\AVE{q}/q_0$ should depend on the specific microstructure, the dependence is
	very weak \cite{Lebenhaft1979}, especially in the regime where $\phi$ is far
	from the closed-packing limit $\phi_{RCP}\approx0.64$.
	The following equation was found to be a universal fit to $\AVE{q}/q_0$ for all
	structures and all $(\Da,\phi)$ ranges investigated in this work
	\begin{equation}\label{eq:qtoq0}
		\frac{\AVE{q}}{q_0}=\frac{1}{1-  B \phi^{1/3}{\Da}/{(1+\Da)}}\, ,
	\end{equation}
	with $B=1.62$.
	
	Although on average particles are always supplied with reactant as the global
	average $\AVE{c}$ is held constant, locally they still compete for reactant due
	to the fluctuations of the microstructure. If the local particle volume fraction
	$\phi$ is higher than the average $\AVE{\phi}$, the competition for reactant
	solutes occurs locally and decreases $q$ of particles with high $\phi_p$. 
	We quantify this local competition by the correlation
	between the single particle reactivity $q$ and the local volume fraction
	$\phi_p$. Fig.~\ref{fig:qphicorr} shows this correlation for an example system
	of $\Da=2,\phi=0.0488$, at different repulsion strengths. 
	Fig.~\ref{fig:qphicorr} A and B both show a power-law correlation between $q/q_0$ and $\phi_p$.
	It is known that the correlation is $q/q_0\propto\phi_p^{-1/2}$ in random suspensions
	but changes to $q/q_0\propto\phi_p^{-1/3}$ in periodic suspensions \cite{Bonnecaze1991}. 
	The simulation results agree with this transition.
	With $S_D=-15$ in Fig.~\ref{fig:qphicorr} A 
	the structure remains random and the correlation follows $q/q_0\propto\phi_p^{-1/2}$.
	Under strong repulsion ($S_D=-60$ in Fig.~\ref{fig:qphicorr} B), the particle reactivity $q$ is
	narrowly distributed around $\AVE{q}$, because the strong repulsion keeps the
	particles almost homogeneously distributed. 
	In this case each particle experiences almost the same microstructure, 
	and the structure is close to a periodic configuration. 
	Thus the correlation in this case is close to $q/q_0\propto\phi_p^{-1/3}$.

	Therefore, in the strong repulsion case, we can ignore the fluctuations in $q$
	and define a parameter based on $\AVE{q}$ to quantify the leading order effect
	of repulsion vs Brownian motion, again by an analogy to an OCP.
	In an OCP, the controlling parameter is $\Gamma_e = (Ze)^2/(4\pi\epsilon_0 L k_B
	T)$, where $Ze$ is the ion charge, $\epsilon_0$ is the dielectric permittivity,
	and $L$ is a length scale determined by ion number density $n$: $L= (4\pi n/3)^{-1/3}$.
	The parameter $\Gamma_e$ measures the ratio of the electrostatic potential energy of two ions
	separated by $L$ to the thermal energy $k_BT$. 
	Similarly, we can define $\Gamma_c$ as the ratio of diffusiophoretic repulsion
	to Brownian motion, where the subscript $c$ denotes chemically active particles.
	To leading order, the repulsive diffusiophoretic velocity $U_0\sim -S_D \nabla
	c$, as shown in (\ref{eq:U0nonD}), and in the over-damped limit $\bF\propto
	6\pi\eta a U_0 \sim \nabla (1/r)$. Thus, we can define an `average potential of chemical force'
	$\varPhi_c$ according to $\bF=-\nabla \varPhi_c$.
	We use the same length scale $L= (4\pi n/3)^{-1/3}$ as in an OCP, but replace the
	number density $n$ with particle volume fraction $\phi$, since particles are not
	point charges.
	We also scale $\AVE{q}$ with $q_0$, as in the scaling relation (\ref{eq:qtoq0}).
	Therefore, we have $\varPhi_L = S_D k_BT \AVE{q}/\left[(\Da+2)\AVE{c}D_R
	L\right]$, and  $\Gamma_c$ can be defined in the nondimensional form:
	\begin{align}\label{eq:GammacDef}
		\Gamma_c = - 4\pi \frac{\Da}{1+\Da} \frac{S_D}{\Da+2} \phi^{1/3}
		\frac{\AVE{q}}{q_0}\,. 
	\end{align}
	Note, the thermal energy $k_BT$ does not appear in $\Gamma_c$ because both the
	repulsive force (in equation~(\ref{eq:U0def})) and thermal Brownian motion scale
	linearly with $k_BT$.
	

	\section{The strong repulsion regime: `liquid-to-crystal' phase transition}
	
	The analogy to an OCP and the similar definition of $\Gamma_c$ implies the
	existence of a liquid-to-crystal phase transition, which is confirmed by our
	simulations.

	In an OCP, BCC is considered the stable crystal structure. However, the free
	energy difference between BCC and FCC is very small, and FCC can also maintain
	its structure, similar to diamond and graphite. The melting point of both FCC
	and BCC are reported \cite{Dubin1990,Stringfellow1990} to be:
	$\Gamma_{e}^{BCC}\approx175$ and $\Gamma_{e}^{FCC}\approx185$, respectively.
	
	For chemically active particles, we conducted  simulations in 3D cubic periodic
	boxes with approximately $N=1000$ particles, with large $S_D$ ($\Gamma_c \sim
	800$), starting from a random particle distribution and tracked the structural
	evolution for a long time $\sim 1000 \tau_D$.
	The simulation process is equivalent to suddenly cooling a liquid to very low
	temperature and allowing it to relax to equilibrium.
	BCC crystals formed in all `cooling' simulations, with inevitable distortion
	and defects.
	The formation of a BCC lattice is similar to the experiments \cite{Tan1995} and
	simulations \cite{Daligault2006} of an OCP. 
	
	In order to accurately locate the transition, i.e., the `melting point' of the
	repulsive active particle crystal, `melting simulations' were conducted.
	Melting, instead of cooling, is chosen because in the liquid-solid phase
	transition the cooling process usually requires a large amount of sub-cooling to
	provide the crystallization with enough `driving force', while melting usually
	occurs immediately at the melting point.
	Although the thermal energy $k_BT$ cancels out in the definition of $\Gamma_c$,
	increasing the Brownian motion is equivalent to increasing the `temperature'
	and corresponds to decreasing $\Gamma_c$. 
	We start from 3D periodic systems of perfect crystal structures and run
	simulations covering a wide range of $\Gamma_c$, for sufficiently long times
	$\sim 1000\tau_D$.
	
	\begin{figure}
		\centering
		\includegraphics[width=\linewidth]{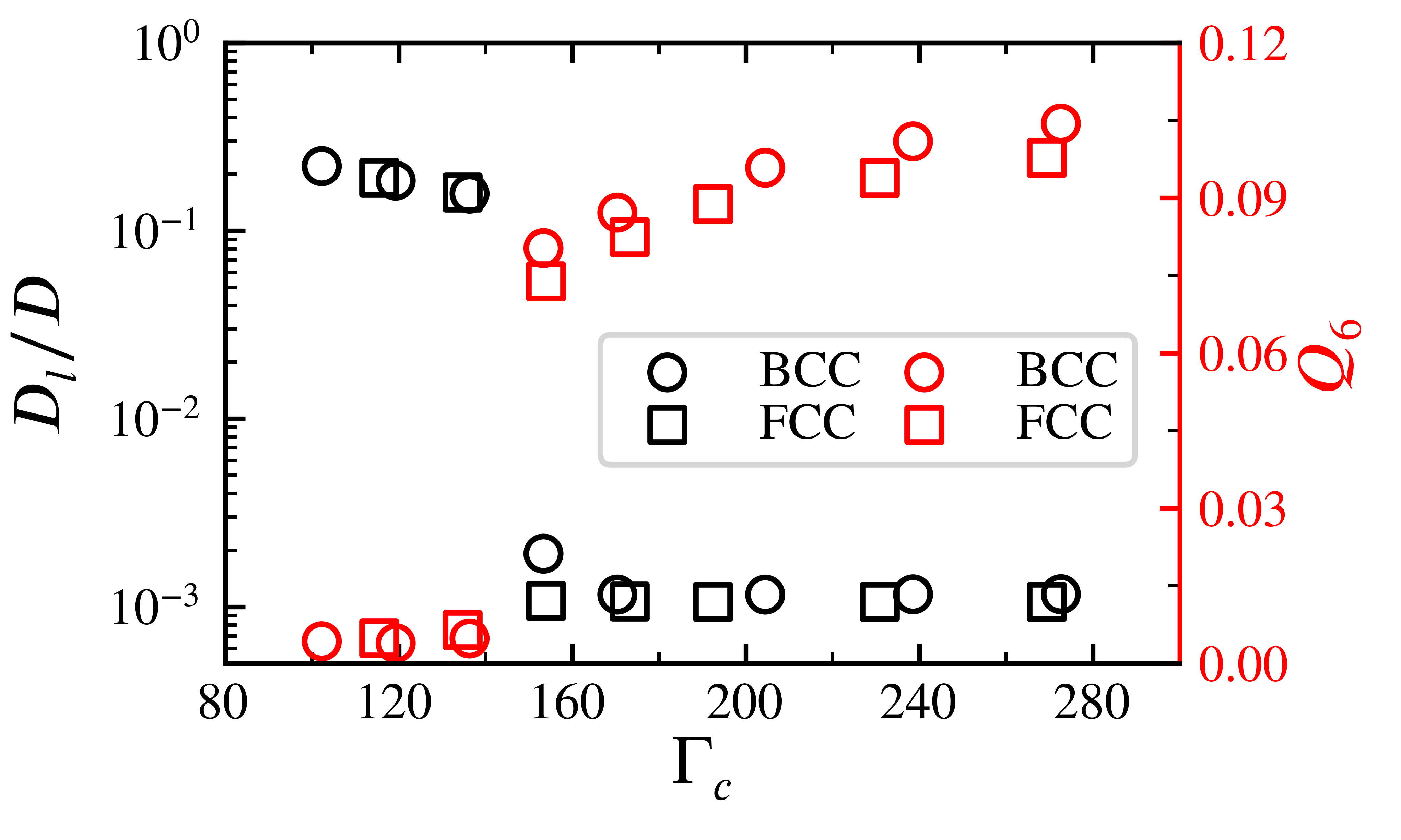}
		\caption{\label{fig:DlD} The measurement of structural change of BCC and FCC
			crystals, both for $\Da=2.0$. $\phi = 0.0388$ for BCC system and $\phi = 0.0488$
			for FCC system.}
	\end{figure}
	
\begin{figure}
		\centering
		\includegraphics[width=\linewidth]{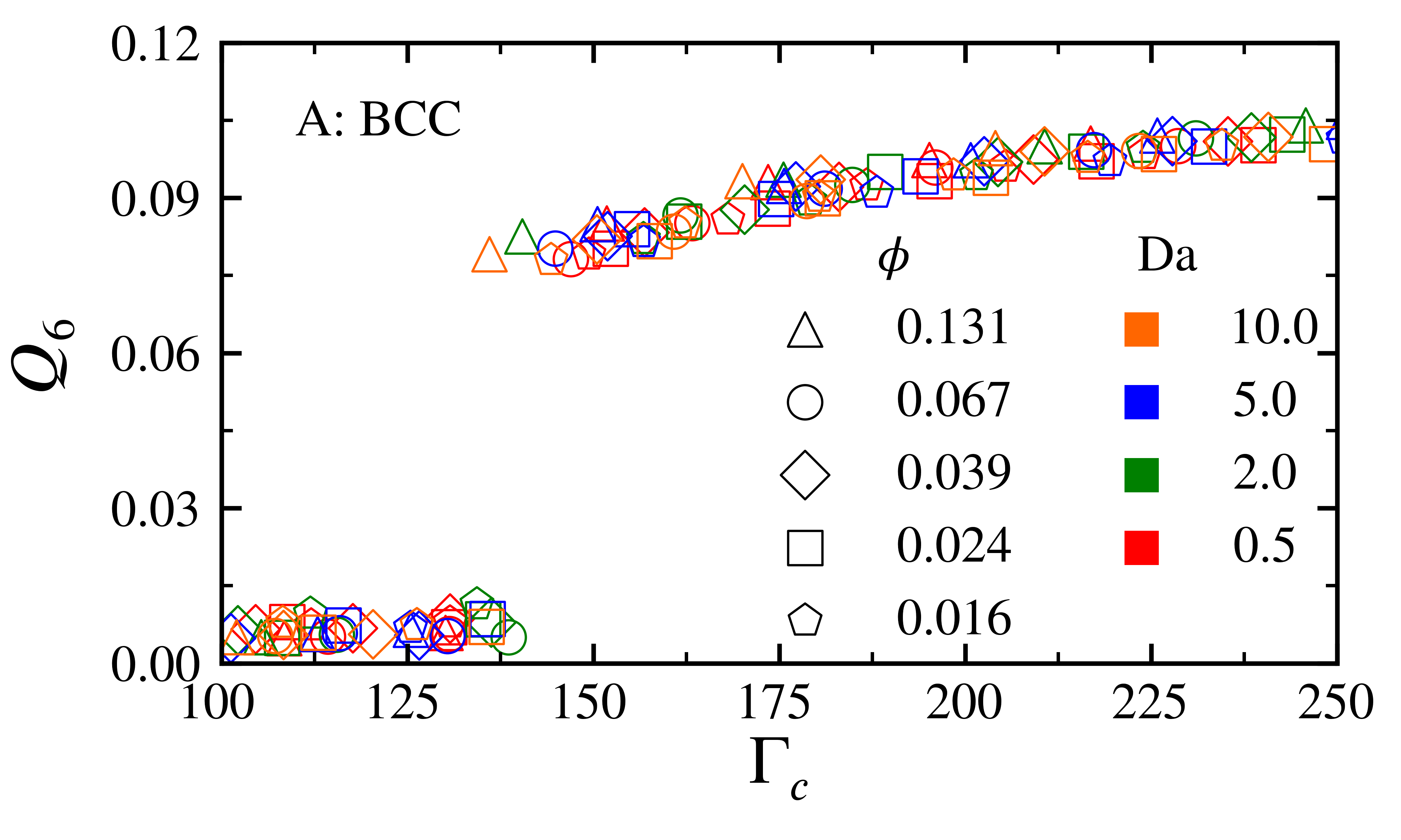}
		\includegraphics[width=\linewidth]{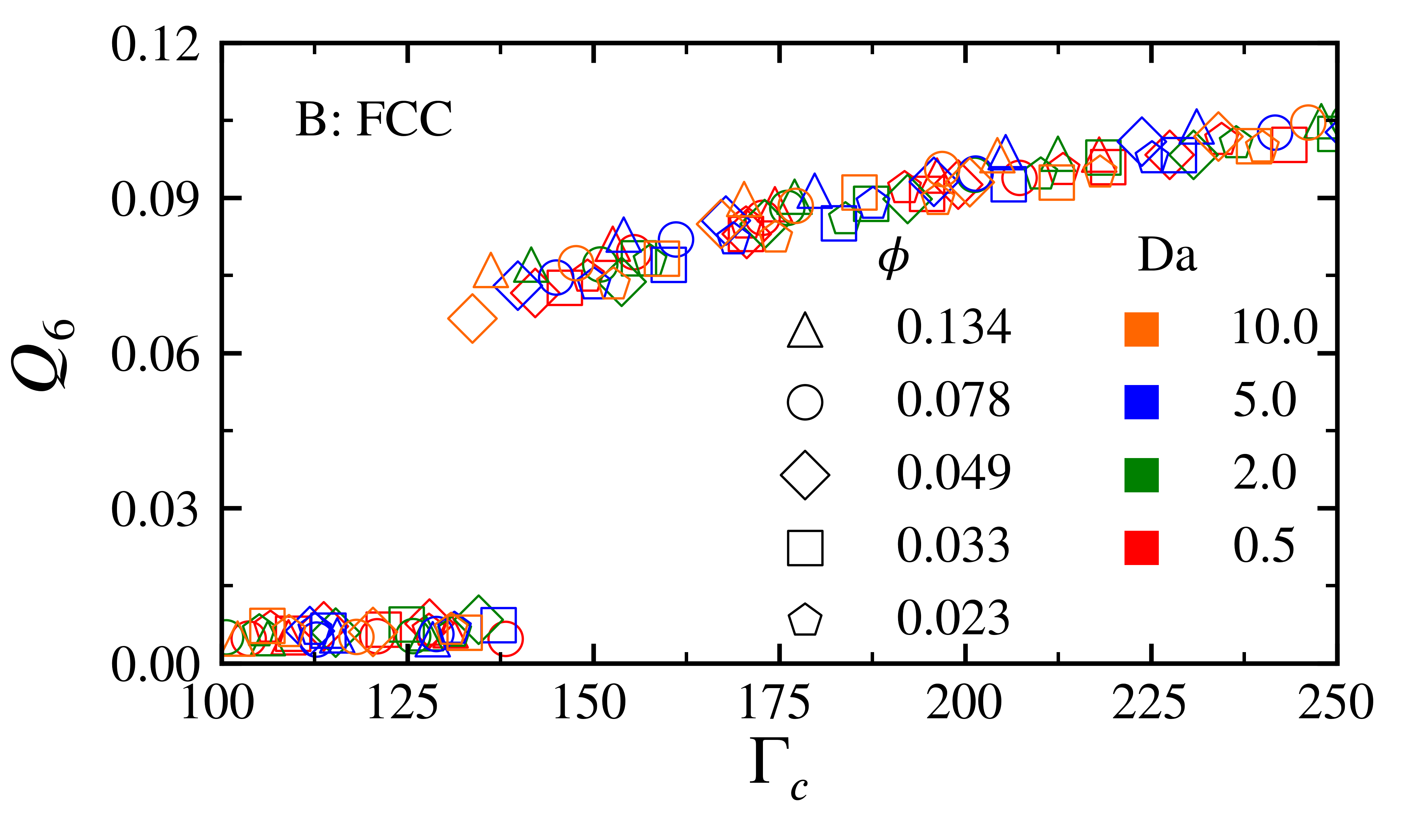}
		\caption{\label{fig:GammaMelt}  The measurement $Q_6$ of the melting process.
			A: The initial configuration is BCC and B: The initial configuration is FCC. For
			each combination of $\phi$ (shape) and $\Da$ (color), simulations of different
			$S_D$ are conducted so that a range of $100<\Gamma_c<250$ is covered. The
			melting point for both BCC and FCC is $\Gamma_c\approx 140$.}
	\end{figure}

	To quantify the structure we use both the dynamic criterion $D_l/D$
	\cite{Lowen1993}, where $D_l$ is the long-time diffusivity of the particles, and
	the static order parameter $Q_6$ \cite{Wolde1995}.  As shown in
	Figs.~\ref{fig:DlD} and \ref{fig:GammaMelt},  $D_l/D$ and $Q_6$ give consistent
	results in quantifying the system `melting point', for both BCC and FCC
	structures. However, the calculation of $D_l/D$ requires significant 
	computation time because we must track the system for a very long time.  Thus,
	we use $Q_6$  when mapping the entire phase diagram for the range of
	$0.001<\phi<0.15$, $0.1<\Da<10$, and $N\approx1000$.
	When calculating $Q_6$ we include approximately the second shell of neighbors 
	\cite{Wolde1995}.   Including only the first shell results in a smaller value
	of $Q_6$, but the measured transition point does not change. 
	Test runs show that a simple cubic lattice spontaneously transforms to a
	distorted BCC lattice. 
	Therefore we search for the melting point of BCC and FCC lattices only.
	
	As shown in Fig.~\ref{fig:GammaMelt}, all the melting simulations show the same
	sharp jump in $Q_6$, and the transition point for both BCC and FCC is
	$\Gamma_{c}^{BCC,FCC} \approx 140$.

	\section{Conclusions \& Discussion}
	We explored repulsive chemically active particles with simulations and showed
	that the system behavior can be determined by a single parameter $\Gamma_c$.
	The `liquid-to-crystal' phase transition is located at
	$\Gamma_{c}^{BCC,FCC}\approx140$, which differs from the OCP results
	$\Gamma_e^{BCC}\approx175,\Gamma_e^{FCC}\approx185$.
	The difference may come from three effects.
	First, although for repulsive chemically active particles the almost homogeneous
	local structure allows us to define $\Gamma_c$ based on $\AVE{q}$,  fluctuations
	in $q$ are still present, which is very different from an OCP system with fixed
	point charges. More importantly, the changing reactivity leads to Brinkman
	screening \cite{Morris1995}, which changes the long-ranged $1/r$ interaction to
	a screened $\exp(r/L_B)/r$, where $L_B \sim a \phi^{-1/2}$ is the screening
	length. The role of screening in repulsive active matter is a complicated issue
	and it is unclear whether it causes the differences in the melting point of
	$\Gamma_c$ compared to $\Gamma_e$.
	Second, limited by computing resources, in simulations we truncated the particle
	multipole expansion at the dipole level,  and so some inaccuracy is inevitable.
	Third, the transition point of an OCP system is typically found by searching for
	the free energy cross-over via Monte-Carlo methods. However, thermodynamics for
	repulsive active particles are not yet defined, and so we have to search for a
	transition point with dynamic simulations, which may give $\sim10\%$ error
	depending on the system property and methodology \cite{Hoffmann2001}.
	
	Regarding the experimental realizations with  hydrogen peroxide and oxygen
	molecules as the fuel, both are at the nanoscale, and in this limit
	$L(\Delta)a^3 \sim\delta^2 a$, where $\delta\sim 10^{-9} \si{\meter}$.
	Therefore, $S_D \sim O(100)$ for $\AVE{c}\sim 1 \si{\mole\per\liter}$, and $\Gamma_c\sim
	O(100)$. So the estimated phase-transition at $\Gamma_c^{BCC,FCC}\sim140$ is
	within the reach of the experiments. If the particles are confined to a
	monolayer by gravity and geometry, similar repulsive crystals should form, which
	should be hexagonal because the repulsion to leading order is isotropic.  It
	would be interesting to see if our predictions are borne out by experiment.
	
	In this work we investigated homogeneously reactive particles. In addition to
	particle-particle interaction, active Janus particle with a reactive hemisphere
	also achieve self-propulsion given by~(\ref{eq:U0def}). For Janus particles no
	repulsive crystal formation is observed in simulations, because Janus particles
	can achieve a much larger velocity, $\bU_0$, so the long-ranged repulsion due to
	diffusiophoresis is not strong enough to trap them in a lattice. Also, it is not
	legitimate to define a $\Gamma_c^J$ by simply replacing the translational
	diffusivity $D=k_BT/\zeta$ with the swim-diffusivity $D^{swim}$ for Janus
	particles and then determine the system dynamics with $\Gamma_c^J$, because
	$D^{swim}$ only appears at a time scale longer than the reorientation time
	$\tau_R$, and the short-time dynamics are important in many cases, such as 
	crystal formation. In fact, it is not clear  whether a meaningful parameter
	$\Gamma_c^J$ can be defined for Janus particles. Moreover, while the swim
	pressure and thermodynamic-like theories \cite{Pressure2014,ThermoAct2015}
	describe attractive active swimmer behaviors well,  it is  not clear whether a
	similar non-equilibrium thermodynamic argument can be conducted to estimate the
	melting point, $\Gamma_c^{BCC,FCC}$, of repulsive swimmers. Repulsive Janus
	particles are left for a future study.

	\begin{acknowledgments}
	We thank Prof. Zhen-Gang Wang for bringing our attention to the repulsive
	system. This work is supported by NSF CBET-1437570.
	\end{acknowledgments}

\bibliography{ref}

\begin{thebibliography}{40}%
\makeatletter
\providecommand \@ifxundefined [1]{%
 \@ifx{#1\undefined}
}%
\providecommand \@ifnum [1]{%
 \ifnum #1\expandafter \@firstoftwo
 \else \expandafter \@secondoftwo
 \fi
}%
\providecommand \@ifx [1]{%
 \ifx #1\expandafter \@firstoftwo
 \else \expandafter \@secondoftwo
 \fi
}%
\providecommand \natexlab [1]{#1}%
\providecommand \enquote  [1]{``#1''}%
\providecommand \bibnamefont  [1]{#1}%
\providecommand \bibfnamefont [1]{#1}%
\providecommand \citenamefont [1]{#1}%
\providecommand \href@noop [0]{\@secondoftwo}%
\providecommand \href [0]{\begingroup \@sanitize@url \@href}%
\providecommand \@href[1]{\@@startlink{#1}\@@href}%
\providecommand \@@href[1]{\endgroup#1\@@endlink}%
\providecommand \@sanitize@url [0]{\catcode `\\12\catcode `\$12\catcode
  `\&12\catcode `\#12\catcode `\^12\catcode `\_12\catcode `\%12\relax}%
\providecommand \@@startlink[1]{}%
\providecommand \@@endlink[0]{}%
\providecommand \url  [0]{\begingroup\@sanitize@url \@url }%
\providecommand \@url [1]{\endgroup\@href {#1}{\urlprefix }}%
\providecommand \urlprefix  [0]{URL }%
\providecommand \Eprint [0]{\href }%
\providecommand \doibase [0]{http://dx.doi.org/}%
\providecommand \selectlanguage [0]{\@gobble}%
\providecommand \bibinfo  [0]{\@secondoftwo}%
\providecommand \bibfield  [0]{\@secondoftwo}%
\providecommand \translation [1]{[#1]}%
\providecommand \BibitemOpen [0]{}%
\providecommand \bibitemStop [0]{}%
\providecommand \bibitemNoStop [0]{.\EOS\space}%
\providecommand \EOS [0]{\spacefactor3000\relax}%
\providecommand \BibitemShut  [1]{\csname bibitem#1\endcsname}%
\let\auto@bib@innerbib\@empty
\bibitem [{\citenamefont {Ebbens}\ and\ \citenamefont
  {Howse}(2010)}]{EbbensReview2010}%
  \BibitemOpen
  \bibfield  {author} {\bibinfo {author} {\bibfnamefont {S.~J.}\ \bibnamefont
  {Ebbens}}\ and\ \bibinfo {author} {\bibfnamefont {J.~R.}\ \bibnamefont
  {Howse}},\ }\href {\doibase 10.1039/B918598D} {\bibfield  {journal} {\bibinfo
   {journal} {Soft Matter}\ }\textbf {\bibinfo {volume} {6}},\ \bibinfo {pages}
  {726} (\bibinfo {year} {2010})}\BibitemShut {NoStop}%
\bibitem [{\citenamefont {Theurkauff}\ \emph {et~al.}(2012)\citenamefont
  {Theurkauff}, \citenamefont {Cottin-Bizonne}, \citenamefont {Palacci},
  \citenamefont {Ybert},\ and\ \citenamefont {Bocquet}}]{Theurkauff2012}%
  \BibitemOpen
  \bibfield  {author} {\bibinfo {author} {\bibfnamefont {I.}~\bibnamefont
  {Theurkauff}}, \bibinfo {author} {\bibfnamefont {C.}~\bibnamefont
  {Cottin-Bizonne}}, \bibinfo {author} {\bibfnamefont {J.}~\bibnamefont
  {Palacci}}, \bibinfo {author} {\bibfnamefont {C.}~\bibnamefont {Ybert}}, \
  and\ \bibinfo {author} {\bibfnamefont {L.}~\bibnamefont {Bocquet}},\ }\href
  {\doibase 10.1103/PhysRevLett.108.268303} {\bibfield  {journal} {\bibinfo
  {journal} {Phys. Rev. Lett.}\ }\textbf {\bibinfo {volume} {108}},\ \bibinfo
  {pages} {268303} (\bibinfo {year} {2012})}\BibitemShut {NoStop}%
\bibitem [{\citenamefont {Buttinoni}\ \emph {et~al.}(2013)\citenamefont
  {Buttinoni}, \citenamefont {Bialk{\'e}}, \citenamefont {K{\"u}mmel},
  \citenamefont {L{\"o}wen}, \citenamefont {Bechinger},\ and\ \citenamefont
  {Speck}}]{Buttinoni2013}%
  \BibitemOpen
  \bibfield  {author} {\bibinfo {author} {\bibfnamefont {I.}~\bibnamefont
  {Buttinoni}}, \bibinfo {author} {\bibfnamefont {J.}~\bibnamefont
  {Bialk{\'e}}}, \bibinfo {author} {\bibfnamefont {F.}~\bibnamefont
  {K{\"u}mmel}}, \bibinfo {author} {\bibfnamefont {H.}~\bibnamefont
  {L{\"o}wen}}, \bibinfo {author} {\bibfnamefont {C.}~\bibnamefont
  {Bechinger}}, \ and\ \bibinfo {author} {\bibfnamefont {T.}~\bibnamefont
  {Speck}},\ }\href {\doibase 10.1103/PhysRevLett.110.238301} {\bibfield
  {journal} {\bibinfo  {journal} {Phys. Rev. Lett.}\ }\textbf {\bibinfo
  {volume} {110}},\ \bibinfo {pages} {238301} (\bibinfo {year}
  {2013})}\BibitemShut {NoStop}%
\bibitem [{\citenamefont {Palacci}\ \emph {et~al.}(2013)\citenamefont
  {Palacci}, \citenamefont {Sacanna}, \citenamefont {Steinberg}, \citenamefont
  {Pine},\ and\ \citenamefont {Chaikin}}]{Palacci2013}%
  \BibitemOpen
  \bibfield  {author} {\bibinfo {author} {\bibfnamefont {J.}~\bibnamefont
  {Palacci}}, \bibinfo {author} {\bibfnamefont {S.}~\bibnamefont {Sacanna}},
  \bibinfo {author} {\bibfnamefont {A.~P.}\ \bibnamefont {Steinberg}}, \bibinfo
  {author} {\bibfnamefont {D.~J.}\ \bibnamefont {Pine}}, \ and\ \bibinfo
  {author} {\bibfnamefont {P.~M.}\ \bibnamefont {Chaikin}},\ }\href {\doibase
  10.1126/science.1230020} {\bibfield  {journal} {\bibinfo  {journal}
  {Science}\ }\textbf {\bibinfo {volume} {339}},\ \bibinfo {pages} {936}
  (\bibinfo {year} {2013})}\BibitemShut {NoStop}%
\bibitem [{\citenamefont {Takatori}\ and\ \citenamefont
  {Brady}(2015)}]{ThermoAct2015}%
  \BibitemOpen
  \bibfield  {author} {\bibinfo {author} {\bibfnamefont {S.~C.}\ \bibnamefont
  {Takatori}}\ and\ \bibinfo {author} {\bibfnamefont {J.~F.}\ \bibnamefont
  {Brady}},\ }\href {\doibase 10.1103/PhysRevE.91.032117} {\bibfield  {journal}
  {\bibinfo  {journal} {Phys. Rev. E}\ }\textbf {\bibinfo {volume} {91}},\
  \bibinfo {pages} {{032117}} (\bibinfo {year} {2015})}\BibitemShut {NoStop}%
\bibitem [{\citenamefont {Takatori}\ \emph {et~al.}(2014)\citenamefont
  {Takatori}, \citenamefont {Yan},\ and\ \citenamefont {Brady}}]{Pressure2014}%
  \BibitemOpen
  \bibfield  {author} {\bibinfo {author} {\bibfnamefont {S.~C.}\ \bibnamefont
  {Takatori}}, \bibinfo {author} {\bibfnamefont {W.}~\bibnamefont {Yan}}, \
  and\ \bibinfo {author} {\bibfnamefont {J.~F.}\ \bibnamefont {Brady}},\ }\href
  {\doibase 10.1103/PhysRevLett.113.028103} {\bibfield  {journal} {\bibinfo
  {journal} {Phys. Rev. Lett.}\ }\textbf {\bibinfo {volume} {113}},\ \bibinfo
  {pages} {{028103}} (\bibinfo {year} {2014})}\BibitemShut {NoStop}%
\bibitem [{\citenamefont {Stenhammar}\ \emph {et~al.}(2013)\citenamefont
  {Stenhammar}, \citenamefont {Tiribocchi}, \citenamefont {Allen},
  \citenamefont {Marenduzzo},\ and\ \citenamefont {Cates}}]{Stenhammar2013}%
  \BibitemOpen
  \bibfield  {author} {\bibinfo {author} {\bibfnamefont {J.}~\bibnamefont
  {Stenhammar}}, \bibinfo {author} {\bibfnamefont {A.}~\bibnamefont
  {Tiribocchi}}, \bibinfo {author} {\bibfnamefont {R.~J.}\ \bibnamefont
  {Allen}}, \bibinfo {author} {\bibfnamefont {D.}~\bibnamefont {Marenduzzo}}, \
  and\ \bibinfo {author} {\bibfnamefont {M.~E.}\ \bibnamefont {Cates}},\ }\href
  {\doibase 10.1103/PhysRevLett.111.145702} {\bibfield  {journal} {\bibinfo
  {journal} {Phys. Rev. Lett.}\ }\textbf {\bibinfo {volume} {111}},\ \bibinfo
  {pages} {145702} (\bibinfo {year} {2013})}\BibitemShut {NoStop}%
\bibitem [{\citenamefont {Solon}\ \emph {et~al.}(2015)\citenamefont {Solon},
  \citenamefont {Fily}, \citenamefont {Baskaran}, \citenamefont {Cates},
  \citenamefont {Kafri}, \citenamefont {Kardar},\ and\ \citenamefont
  {Tailleur}}]{Solon2015a}%
  \BibitemOpen
  \bibfield  {author} {\bibinfo {author} {\bibfnamefont {A.~P.}\ \bibnamefont
  {Solon}}, \bibinfo {author} {\bibfnamefont {Y.}~\bibnamefont {Fily}},
  \bibinfo {author} {\bibfnamefont {A.}~\bibnamefont {Baskaran}}, \bibinfo
  {author} {\bibfnamefont {M.~E.}\ \bibnamefont {Cates}}, \bibinfo {author}
  {\bibfnamefont {Y.}~\bibnamefont {Kafri}}, \bibinfo {author} {\bibfnamefont
  {M.}~\bibnamefont {Kardar}}, \ and\ \bibinfo {author} {\bibfnamefont
  {J.}~\bibnamefont {Tailleur}},\ }\href {\doibase 10.1038/nphys3377}
  {\bibfield  {journal} {\bibinfo  {journal} {Nat. Phys.}\ }\textbf {\bibinfo
  {volume} {11}},\ \bibinfo {pages} {673} (\bibinfo {year} {2015})}\BibitemShut
  {NoStop}%
\bibitem [{\citenamefont {Cates}\ and\ \citenamefont
  {Tailleur}(2015)}]{Cates2015}%
  \BibitemOpen
  \bibfield  {author} {\bibinfo {author} {\bibfnamefont {M.~E.}\ \bibnamefont
  {Cates}}\ and\ \bibinfo {author} {\bibfnamefont {J.}~\bibnamefont
  {Tailleur}},\ }\href {\doibase 10.1146/annurev-conmatphys-031214-014710}
  {\bibfield  {journal} {\bibinfo  {journal} {Annu. Rev. Condens. Matter
  Phys.}\ }\textbf {\bibinfo {volume} {6}},\ \bibinfo {pages} {219} (\bibinfo
  {year} {2015})}\BibitemShut {NoStop}%
\bibitem [{\citenamefont {Soh}\ \emph {et~al.}(2008)\citenamefont {Soh},
  \citenamefont {Bishop},\ and\ \citenamefont {Grzybowski}}]{Soh2008}%
  \BibitemOpen
  \bibfield  {author} {\bibinfo {author} {\bibfnamefont {S.}~\bibnamefont
  {Soh}}, \bibinfo {author} {\bibfnamefont {K.~J.~M.}\ \bibnamefont {Bishop}},
  \ and\ \bibinfo {author} {\bibfnamefont {B.~A.}\ \bibnamefont {Grzybowski}},\
  }\href {\doibase 10.1021/jp7111457} {\bibfield  {journal} {\bibinfo
  {journal} {J. Phys. Chem. B}\ }\textbf {\bibinfo {volume} {112}},\ \bibinfo
  {pages} {10848} (\bibinfo {year} {2008})}\BibitemShut {NoStop}%
\bibitem [{\citenamefont {Derjaguin}\ and\ \citenamefont
  {Golovanov}(1984)}]{Derjaguin1984}%
  \BibitemOpen
  \bibfield  {author} {\bibinfo {author} {\bibfnamefont {B.}~\bibnamefont
  {Derjaguin}}\ and\ \bibinfo {author} {\bibfnamefont {M.}~\bibnamefont
  {Golovanov}},\ }\href {\doibase 10.1016/0166-6622(84)80009-8} {\bibfield
  {journal} {\bibinfo  {journal} {Colloids and Surfaces}\ }\textbf {\bibinfo
  {volume} {10}},\ \bibinfo {pages} {77} (\bibinfo {year} {1984})}\BibitemShut
  {NoStop}%
\bibitem [{\citenamefont {Brush}\ \emph {et~al.}(1966)\citenamefont {Brush},
  \citenamefont {Sahlin},\ and\ \citenamefont {Teller}}]{Brush1966}%
  \BibitemOpen
  \bibfield  {author} {\bibinfo {author} {\bibfnamefont {S.~G.}\ \bibnamefont
  {Brush}}, \bibinfo {author} {\bibfnamefont {H.~L.}\ \bibnamefont {Sahlin}}, \
  and\ \bibinfo {author} {\bibfnamefont {E.}~\bibnamefont {Teller}},\ }\href
  {\doibase 10.1063/1.1727895} {\bibfield  {journal} {\bibinfo  {journal} {J.
  Chem. Phys.}\ }\textbf {\bibinfo {volume} {45}},\ \bibinfo {pages} {2102}
  (\bibinfo {year} {1966})}\BibitemShut {NoStop}%
\bibitem [{\citenamefont {Gillan}(1974)}]{Gillan1974}%
  \BibitemOpen
  \bibfield  {author} {\bibinfo {author} {\bibfnamefont {M.~J.}\ \bibnamefont
  {Gillan}},\ }\href {\doibase 10.1088/0022-3719/7/1/001} {\bibfield  {journal}
  {\bibinfo  {journal} {J. Phys. C}\ }\textbf {\bibinfo {volume} {7}},\
  \bibinfo {pages} {L1} (\bibinfo {year} {1974})}\BibitemShut {NoStop}%
\bibitem [{\citenamefont {Rogers}(1974)}]{Rogers1974}%
  \BibitemOpen
  \bibfield  {author} {\bibinfo {author} {\bibfnamefont {F.~J.}\ \bibnamefont
  {Rogers}},\ }\href {\doibase 10.1103/PhysRevA.10.2441} {\bibfield  {journal}
  {\bibinfo  {journal} {Phys. Rev. A}\ }\textbf {\bibinfo {volume} {10}},\
  \bibinfo {pages} {2441} (\bibinfo {year} {1974})}\BibitemShut {NoStop}%
\bibitem [{\citenamefont {Stroud}\ and\ \citenamefont
  {Ashcroft}(1976)}]{Stroud1976}%
  \BibitemOpen
  \bibfield  {author} {\bibinfo {author} {\bibfnamefont {D.}~\bibnamefont
  {Stroud}}\ and\ \bibinfo {author} {\bibfnamefont {N.}~\bibnamefont
  {Ashcroft}},\ }\href {\doibase 10.1103/PhysRevA.13.1660} {\bibfield
  {journal} {\bibinfo  {journal} {Phys. Rev. A}\ }\textbf {\bibinfo {volume}
  {13}},\ \bibinfo {pages} {1660} (\bibinfo {year} {1976})}\BibitemShut
  {NoStop}%
\bibitem [{\citenamefont {Itoh}\ and\ \citenamefont
  {Ichimaru}(1977)}]{Itoh1977}%
  \BibitemOpen
  \bibfield  {author} {\bibinfo {author} {\bibfnamefont {N.}~\bibnamefont
  {Itoh}}\ and\ \bibinfo {author} {\bibfnamefont {S.}~\bibnamefont
  {Ichimaru}},\ }\href {\doibase 10.1103/PhysRevA.16.2178} {\bibfield
  {journal} {\bibinfo  {journal} {Phys. Rev. A}\ }\textbf {\bibinfo {volume}
  {16}},\ \bibinfo {pages} {2178} (\bibinfo {year} {1977})}\BibitemShut
  {NoStop}%
\bibitem [{\citenamefont {Bernu}(1979)}]{Bernu1979}%
  \BibitemOpen
  \bibfield  {author} {\bibinfo {author} {\bibfnamefont {B.}~\bibnamefont
  {Bernu}},\ }\href {\doibase 10.1007/Bf01009611} {\bibfield  {journal}
  {\bibinfo  {journal} {J. Stat. Phys.}\ }\textbf {\bibinfo {volume} {21}},\
  \bibinfo {pages} {447} (\bibinfo {year} {1979})}\BibitemShut {NoStop}%
\bibitem [{\citenamefont {Baus}\ and\ \citenamefont {Hansen}(1980)}]{Baus1980}%
  \BibitemOpen
  \bibfield  {author} {\bibinfo {author} {\bibfnamefont {M.}~\bibnamefont
  {Baus}}\ and\ \bibinfo {author} {\bibfnamefont {J.~P.}\ \bibnamefont
  {Hansen}},\ }\href {\doibase 10.1016/0370-1573(80)90022-8} {\bibfield
  {journal} {\bibinfo  {journal} {Phys. Rep.}\ }\textbf {\bibinfo {volume}
  {59}},\ \bibinfo {pages} {1} (\bibinfo {year} {1980})}\BibitemShut {NoStop}%
\bibitem [{\citenamefont {Tan}\ \emph {et~al.}(1995)\citenamefont {Tan},
  \citenamefont {Bollinger}, \citenamefont {Jelenkovic},\ and\ \citenamefont
  {Wineland}}]{Tan1995}%
  \BibitemOpen
  \bibfield  {author} {\bibinfo {author} {\bibfnamefont {J.~N.}\ \bibnamefont
  {Tan}}, \bibinfo {author} {\bibfnamefont {J.~J.}\ \bibnamefont {Bollinger}},
  \bibinfo {author} {\bibfnamefont {B.}~\bibnamefont {Jelenkovic}}, \ and\
  \bibinfo {author} {\bibfnamefont {D.~J.}\ \bibnamefont {Wineland}},\ }\href
  {\doibase 10.1103/PhysRevLett.75.4198} {\bibfield  {journal} {\bibinfo
  {journal} {Phys. Rev. Lett.}\ }\textbf {\bibinfo {volume} {75}},\ \bibinfo
  {pages} {4198} (\bibinfo {year} {1995})}\BibitemShut {NoStop}%
\bibitem [{\citenamefont {DeWitt}\ \emph {et~al.}(2001)\citenamefont {DeWitt},
  \citenamefont {Slattery}, \citenamefont {Baiko},\ and\ \citenamefont
  {Yakovlev}}]{DeWitt2001}%
  \BibitemOpen
  \bibfield  {author} {\bibinfo {author} {\bibfnamefont {H.}~\bibnamefont
  {DeWitt}}, \bibinfo {author} {\bibfnamefont {W.}~\bibnamefont {Slattery}},
  \bibinfo {author} {\bibfnamefont {D.}~\bibnamefont {Baiko}}, \ and\ \bibinfo
  {author} {\bibfnamefont {D.}~\bibnamefont {Yakovlev}},\ }\href {\doibase
  10.1002/1521-3986(200103)41:2/3<251::AID-CTPP251>3.0.CO;2-G} {\bibfield
  {journal} {\bibinfo  {journal} {Contrib. Plasma Phys.}\ }\textbf {\bibinfo
  {volume} {41}},\ \bibinfo {pages} {251} (\bibinfo {year} {2001})}\BibitemShut
  {NoStop}%
\bibitem [{\citenamefont {Chugunov}\ \emph {et~al.}(2003)\citenamefont
  {Chugunov}, \citenamefont {Baiko}, \citenamefont {Yakovlev}, \citenamefont
  {{De Witt}},\ and\ \citenamefont {Slattery}}]{Chugunov2003}%
  \BibitemOpen
  \bibfield  {author} {\bibinfo {author} {\bibfnamefont {A.~I.}\ \bibnamefont
  {Chugunov}}, \bibinfo {author} {\bibfnamefont {D.~A.}\ \bibnamefont {Baiko}},
  \bibinfo {author} {\bibfnamefont {D.~G.}\ \bibnamefont {Yakovlev}}, \bibinfo
  {author} {\bibfnamefont {H.~E.}\ \bibnamefont {{De Witt}}}, \ and\ \bibinfo
  {author} {\bibfnamefont {W.~L.}\ \bibnamefont {Slattery}},\ }\href {\doibase
  10.1016/s0378-4371(02)02027-7} {\bibfield  {journal} {\bibinfo  {journal}
  {Physica A}\ }\textbf {\bibinfo {volume} {323}},\ \bibinfo {pages} {413}
  (\bibinfo {year} {2003})}\BibitemShut {NoStop}%
\bibitem [{\citenamefont {Daligault}(2006)}]{Daligault2006}%
  \BibitemOpen
  \bibfield  {author} {\bibinfo {author} {\bibfnamefont {J.}~\bibnamefont
  {Daligault}},\ }\href {\doibase 10.1103/PhysRevE.73.056407} {\bibfield
  {journal} {\bibinfo  {journal} {Phys. Rev. E}\ }\textbf {\bibinfo {volume}
  {73}},\ \bibinfo {pages} {056407} (\bibinfo {year} {2006})}\BibitemShut
  {NoStop}%
\bibitem [{\citenamefont {Yan}\ and\ \citenamefont
  {Brady}(2016)}]{yan_method_2016}%
  \BibitemOpen
  \bibfield  {author} {\bibinfo {author} {\bibfnamefont {W.}~\bibnamefont
  {Yan}}\ and\ \bibinfo {author} {\bibfnamefont {J.~F.}\ \bibnamefont
  {Brady}},\ }\href {\doibase 10.1063/1.4963722} {\bibfield  {journal}
  {\bibinfo  {journal} {The Journal of Chemical Physics}\ }\textbf {\bibinfo
  {volume} {145}},\ \bibinfo {pages} {134902} (\bibinfo {year}
  {2016})}\BibitemShut {NoStop}%
\bibitem [{\citenamefont {C\'{o}rdova-Figueroa}\ and\ \citenamefont
  {Brady}(2008)}]{CordovaFigueroa2008}%
  \BibitemOpen
  \bibfield  {author} {\bibinfo {author} {\bibfnamefont {U.~M.}\ \bibnamefont
  {C\'{o}rdova-Figueroa}}\ and\ \bibinfo {author} {\bibfnamefont {J.~F.}\
  \bibnamefont {Brady}},\ }\href {\doibase 10.1103/PhysRevLett.100.158303}
  {\bibfield  {journal} {\bibinfo  {journal} {Phys. Rev. Lett.}\ }\textbf
  {\bibinfo {volume} {100}},\ \bibinfo {pages} {158303} (\bibinfo {year}
  {2008})}\BibitemShut {NoStop}%
\bibitem [{\citenamefont {Brady}(2011)}]{Brady2011}%
  \BibitemOpen
  \bibfield  {author} {\bibinfo {author} {\bibfnamefont {J.~F.}\ \bibnamefont
  {Brady}},\ }\href {\doibase 10.1017/S0022112010004404} {\bibfield  {journal}
  {\bibinfo  {journal} {J. Fluid Mech.}\ }\textbf {\bibinfo {volume} {667}},\
  \bibinfo {pages} {216} (\bibinfo {year} {2011})}\BibitemShut {NoStop}%
\bibitem [{\citenamefont {Bonnecaze}\ and\ \citenamefont
  {Brady}(1991{\natexlab{a}})}]{Bonnecaze1991a}%
  \BibitemOpen
  \bibfield  {author} {\bibinfo {author} {\bibfnamefont {R.~T.}\ \bibnamefont
  {Bonnecaze}}\ and\ \bibinfo {author} {\bibfnamefont {J.~F.}\ \bibnamefont
  {Brady}},\ }\href {\doibase 10.1098/rspa.1991.0025} {\bibfield  {journal}
  {\bibinfo  {journal} {P. Roy. Soc. Lond. A Mat.}\ }\textbf {\bibinfo {volume}
  {432}},\ \bibinfo {pages} {445} (\bibinfo {year}
  {1991}{\natexlab{a}})}\BibitemShut {NoStop}%
\bibitem [{\citenamefont {Redner}\ \emph {et~al.}(2013)\citenamefont {Redner},
  \citenamefont {Baskaran},\ and\ \citenamefont {Hagan}}]{Redner2013a}%
  \BibitemOpen
  \bibfield  {author} {\bibinfo {author} {\bibfnamefont {G.~S.}\ \bibnamefont
  {Redner}}, \bibinfo {author} {\bibfnamefont {A.}~\bibnamefont {Baskaran}}, \
  and\ \bibinfo {author} {\bibfnamefont {M.~F.}\ \bibnamefont {Hagan}},\ }\href
  {\doibase 10.1103/PhysRevE.88.012305} {\bibfield  {journal} {\bibinfo
  {journal} {Phys. Rev. E}\ }\textbf {\bibinfo {volume} {88}},\ \bibinfo
  {pages} {012305} (\bibinfo {year} {2013})}\BibitemShut {NoStop}%
\bibitem [{\citenamefont {Louis}(2002)}]{Louis2002}%
  \BibitemOpen
  \bibfield  {author} {\bibinfo {author} {\bibfnamefont {A.~A.}\ \bibnamefont
  {Louis}},\ }\href {\doibase 10.1088/0953-8984/14/40/311} {\bibfield
  {journal} {\bibinfo  {journal} {J. Phys. Condens. Matter}\ }\textbf {\bibinfo
  {volume} {14}},\ \bibinfo {pages} {9187} (\bibinfo {year}
  {2002})}\BibitemShut {NoStop}%
\bibitem [{\citenamefont {Tejero}\ and\ \citenamefont
  {Baus}(2003)}]{Tejero2003}%
  \BibitemOpen
  \bibfield  {author} {\bibinfo {author} {\bibfnamefont {C.~F.}\ \bibnamefont
  {Tejero}}\ and\ \bibinfo {author} {\bibfnamefont {M.}~\bibnamefont {Baus}},\
  }\href {\doibase 10.1063/1.1526837} {\bibfield  {journal} {\bibinfo
  {journal} {J. Chem. Phys.}\ }\textbf {\bibinfo {volume} {118}},\ \bibinfo
  {pages} {892} (\bibinfo {year} {2003})}\BibitemShut {NoStop}%
\bibitem [{\citenamefont {Foss}\ and\ \citenamefont {Brady}(2000)}]{Foss2000}%
  \BibitemOpen
  \bibfield  {author} {\bibinfo {author} {\bibfnamefont {D.~R.}\ \bibnamefont
  {Foss}}\ and\ \bibinfo {author} {\bibfnamefont {J.~F.}\ \bibnamefont
  {Brady}},\ }\href {\doibase 10.1122/1.551104} {\bibfield  {journal} {\bibinfo
   {journal} {J. Rheol.}\ }\textbf {\bibinfo {volume} {44}},\ \bibinfo {pages}
  {629} (\bibinfo {year} {2000})}\BibitemShut {NoStop}%
\bibitem [{\citenamefont {Howse}\ \emph {et~al.}(2007)\citenamefont {Howse},
  \citenamefont {Jones}, \citenamefont {Ryan}, \citenamefont {Gough},
  \citenamefont {Vafabakhsh},\ and\ \citenamefont {Golestanian}}]{Howse2007}%
  \BibitemOpen
  \bibfield  {author} {\bibinfo {author} {\bibfnamefont {J.~R.}\ \bibnamefont
  {Howse}}, \bibinfo {author} {\bibfnamefont {R.~A.~L.}\ \bibnamefont {Jones}},
  \bibinfo {author} {\bibfnamefont {A.~J.}\ \bibnamefont {Ryan}}, \bibinfo
  {author} {\bibfnamefont {T.}~\bibnamefont {Gough}}, \bibinfo {author}
  {\bibfnamefont {R.}~\bibnamefont {Vafabakhsh}}, \ and\ \bibinfo {author}
  {\bibfnamefont {R.}~\bibnamefont {Golestanian}},\ }\href {\doibase
  10.1103/PhysRevLett.99.048102} {\bibfield  {journal} {\bibinfo  {journal}
  {Phys. Rev. Lett.}\ }\textbf {\bibinfo {volume} {99}},\ \bibinfo {pages}
  {048102} (\bibinfo {year} {2007})}\BibitemShut {NoStop}%
\bibitem [{\citenamefont {Yan}(2016)}]{yan_dynamics_2016}%
  \BibitemOpen
  \bibfield  {author} {\bibinfo {author} {\bibfnamefont {W.}~\bibnamefont
  {Yan}},\ }\emph {\bibinfo {title} {Dynamics of chemically active
  suspensions}},\ \href
  {http://resolver.caltech.edu/CaltechTHESIS:05242016-214836974} {\bibinfo
  {type} {{PhD} {Thesis}}},\ \bibinfo  {school} {California Institute of
  Technology} (\bibinfo {year} {2016})\BibitemShut {NoStop}%
\bibitem [{\citenamefont {Lebenhaft}\ and\ \citenamefont
  {Kapral}(1979)}]{Lebenhaft1979}%
  \BibitemOpen
  \bibfield  {author} {\bibinfo {author} {\bibfnamefont {J.~R.}\ \bibnamefont
  {Lebenhaft}}\ and\ \bibinfo {author} {\bibfnamefont {R.}~\bibnamefont
  {Kapral}},\ }\href {\doibase 10.1007/BF01013745} {\bibfield  {journal}
  {\bibinfo  {journal} {J. Stat. Phys.}\ }\textbf {\bibinfo {volume} {20}},\
  \bibinfo {pages} {25} (\bibinfo {year} {1979})}\BibitemShut {NoStop}%
\bibitem [{\citenamefont {Bonnecaze}\ and\ \citenamefont
  {Brady}(1991{\natexlab{b}})}]{Bonnecaze1991}%
  \BibitemOpen
  \bibfield  {author} {\bibinfo {author} {\bibfnamefont {R.~T.}\ \bibnamefont
  {Bonnecaze}}\ and\ \bibinfo {author} {\bibfnamefont {J.~F.}\ \bibnamefont
  {Brady}},\ }\href {\doibase 10.1063/1.460372} {\bibfield  {journal} {\bibinfo
   {journal} {J. Chem. Phys.}\ }\textbf {\bibinfo {volume} {94}},\ \bibinfo
  {pages} {537} (\bibinfo {year} {1991}{\natexlab{b}})}\BibitemShut {NoStop}%
\bibitem [{\citenamefont {Dubin}(1990)}]{Dubin1990}%
  \BibitemOpen
  \bibfield  {author} {\bibinfo {author} {\bibfnamefont {D.~H.~E.}\
  \bibnamefont {Dubin}},\ }\href {\doibase 10.1103/PhysRevA.42.4972} {\bibfield
   {journal} {\bibinfo  {journal} {Phys. Rev. A}\ }\textbf {\bibinfo {volume}
  {42}},\ \bibinfo {pages} {4972} (\bibinfo {year} {1990})}\BibitemShut
  {NoStop}%
\bibitem [{\citenamefont {Stringfellow}\ \emph {et~al.}(1990)\citenamefont
  {Stringfellow}, \citenamefont {DeWitt},\ and\ \citenamefont
  {Slattery}}]{Stringfellow1990}%
  \BibitemOpen
  \bibfield  {author} {\bibinfo {author} {\bibfnamefont {G.~S.}\ \bibnamefont
  {Stringfellow}}, \bibinfo {author} {\bibfnamefont {H.~E.}\ \bibnamefont
  {DeWitt}}, \ and\ \bibinfo {author} {\bibfnamefont {W.~L.}\ \bibnamefont
  {Slattery}},\ }\href {\doibase 10.1103/PhysRevA.41.1105} {\bibfield
  {journal} {\bibinfo  {journal} {Phys. Rev. A}\ }\textbf {\bibinfo {volume}
  {41}},\ \bibinfo {pages} {1105} (\bibinfo {year} {1990})}\BibitemShut
  {NoStop}%
\bibitem [{\citenamefont {L{\"o}wen}\ \emph {et~al.}(1993)\citenamefont
  {L{\"o}wen}, \citenamefont {Palberg},\ and\ \citenamefont
  {Simon}}]{Lowen1993}%
  \BibitemOpen
  \bibfield  {author} {\bibinfo {author} {\bibfnamefont {H.}~\bibnamefont
  {L{\"o}wen}}, \bibinfo {author} {\bibfnamefont {T.}~\bibnamefont {Palberg}},
  \ and\ \bibinfo {author} {\bibfnamefont {R.}~\bibnamefont {Simon}},\ }\href
  {\doibase 10.1103/PhysRevLett.70.1557} {\bibfield  {journal} {\bibinfo
  {journal} {Phys. Rev. Lett.}\ }\textbf {\bibinfo {volume} {70}},\ \bibinfo
  {pages} {1557} (\bibinfo {year} {1993})}\BibitemShut {NoStop}%
\bibitem [{\citenamefont {{ten Wolde}}\ \emph {et~al.}(1995)\citenamefont {{ten
  Wolde}}, \citenamefont {Ruiz-Montero},\ and\ \citenamefont
  {Frenkel}}]{Wolde1995}%
  \BibitemOpen
  \bibfield  {author} {\bibinfo {author} {\bibfnamefont {P.~R.}\ \bibnamefont
  {{ten Wolde}}}, \bibinfo {author} {\bibfnamefont {M.~J.}\ \bibnamefont
  {Ruiz-Montero}}, \ and\ \bibinfo {author} {\bibfnamefont {D.}~\bibnamefont
  {Frenkel}},\ }\href {\doibase 10.1103/PhysRevLett.75.2714} {\bibfield
  {journal} {\bibinfo  {journal} {Phys. Rev. Lett.}\ }\textbf {\bibinfo
  {volume} {75}},\ \bibinfo {pages} {2714} (\bibinfo {year}
  {1995})}\BibitemShut {NoStop}%
\bibitem [{\citenamefont {Morris}\ and\ \citenamefont
  {Brady}(1995)}]{Morris1995}%
  \BibitemOpen
  \bibfield  {author} {\bibinfo {author} {\bibfnamefont {J.~F.}\ \bibnamefont
  {Morris}}\ and\ \bibinfo {author} {\bibfnamefont {J.~F.}\ \bibnamefont
  {Brady}},\ }\href {\doibase 10.1021/ie00037a040} {\bibfield  {journal}
  {\bibinfo  {journal} {Ind. Eng. Chem. Res.}\ }\textbf {\bibinfo {volume}
  {34}},\ \bibinfo {pages} {3514} (\bibinfo {year} {1995})}\BibitemShut
  {NoStop}%
\bibitem [{\citenamefont {Hoffmann}\ and\ \citenamefont
  {L{\"o}wen}(2001)}]{Hoffmann2001}%
  \BibitemOpen
  \bibfield  {author} {\bibinfo {author} {\bibfnamefont {G.~P.}\ \bibnamefont
  {Hoffmann}}\ and\ \bibinfo {author} {\bibfnamefont {H.}~\bibnamefont
  {L{\"o}wen}},\ }\href {\doibase 10.1088/0953-8984/13/41/311} {\bibfield
  {journal} {\bibinfo  {journal} {J. Phys. Condens. Matter}\ }\textbf {\bibinfo
  {volume} {13}},\ \bibinfo {pages} {9197} (\bibinfo {year}
  {2001})}\BibitemShut {NoStop}%
\end{thebibliography}

\end{document}